\documentclass[journal]{IEEEtran}

\usepackage{amsmath,graphicx,array,subfigure,nccbbb,url,bm}
\usepackage{multirow}
\usepackage{amssymb}

\usepackage{amsthm}
\newtheorem{prop}{Proposition}
\theoremstyle{definition}
\newtheorem{prof}{Proof}

\makeatletter

\newcommand{\Rmnum}[1]{\expandafter\@slowromancap\romannumeral #1@}

\usepackage{comment}
\usepackage{algorithmicx}
\usepackage{algorithm}
\usepackage{algpseudocode}

\usepackage{booktabs} 

\usepackage{xcolor}

\makeatother

\begin{document}

\title{Deep Adversarial Learning in Intrusion Detection: A Data Augmentation Enhanced Framework}

\author{He Zhang,~\IEEEmembership{}
        Xingrui Yu,~\IEEEmembership{}
        Peng Ren,~\IEEEmembership{Senior Member, IEEE}\\
        Chunbo Luo,~\IEEEmembership{Member, IEEE}
        and~Geyong Min,~\IEEEmembership{Senior Member, IEEE}
\thanks{H. Zhang, C. Luo, and G. Min are with the College of Engineering, Mathematics and Physical Sciences, University of Exeter, Exeter EX4 4QF, UK (e-mail: hz298@exeter.ac.uk; c.luo@exeter.ac.uk; g.min@exeter.ac.uk).

X. Yu, and P. Ren are with the College of Information and Control Engineering, China University of Petroleum (East China), Qingdao 266580, China (email: xingrui\_yu@hotmail.com; pengren@upc.edu.cn).}
}


\markboth{}%
{ZHANG \MakeLowercase{\textit{et al.}}: Deep Adversarial Learning in Intrusion Detection: A Data Augmentation Enhanced Framework}

\maketitle

\begin{abstract}
  Intrusion detection systems (IDSs) play an important role in identifying malicious attacks and threats in networking systems. As  fundamental tools of IDSs, learning based classification methods have been widely employed. When it comes to detecting network intrusions in small sample sizes (e.g., emerging intrusions), the limited number and imbalanced proportion of training samples usually cause significant challenges in training supervised and semi-supervised classifiers. In this paper, we propose a general network intrusion detection framework to address the challenges of both \emph{data scarcity} and \emph{data imbalance}. The novelty of the proposed framework focuses on incorporating deep adversarial learning with statistical learning and exploiting learning based data augmentation. Given a small set of network intrusion samples, it first derives a Poisson-Gamma joint probabilistic generative model to generate synthesised intrusion data using Monte Carlo methods. Those synthesised data are then augmented by deep generative neural networks through adversarial learning. Finally, it adopts the augmented intrusion data to train supervised models for detecting network intrusions. Comprehensive experimental validations on KDD Cup 99 dataset show that the proposed framework outperforms the existing learning based IDSs in terms of improved accuracy, precision, recall, and F1-score.
\end{abstract}


\IEEEpeerreviewmaketitle

\section{Introduction}

\IEEEPARstart{N}{etwork} intrusion detection (NID) has been one of the key components towards network security in the past decades \cite{10-NIDReview}, \cite{16BICT-DL4NIDS}, \cite{18-NIDvDL}. Intrusion detection systems (IDSs) play a significant role in NID for deterring cyber attacks and threats coming from a vast number of network-connected devices (e.g., computing systems \cite{17Anomaly-Detection} \cite{18Anomaly-Detection}, sensor networks \cite{12TON-NID4MSN} \cite{14TIFS-NIDWSN} and software defined networks \cite{16-NIDvDL4SDN}). With the increasing usage of networking systems and frequent emergence of varied intrusion attacks, new generation of high performance IDSs needs to be developed for accurately detecting not only a series of high-frequency network intrusions, such as \verb"neptune" and \verb"smurf" attacks, but also intrusions (e.g., \verb"mailbomb" and \verb"snmpguess" attacks) that have limited known records.

IDSs can be generally categorized into two categories. The first group focus on patterns/signatures of network packets/traffic. Those systems identify network intrusions using rule-based matching \cite{14TON-NIDvRule4Pack}, \cite{16TON-NIDvMatch4Pack}. The second group apply machine learning (ML) based approaches such as supervised and/or semi-supervised learning and train NID models on a collection of labeled and/or unlabeled network data. Those learning based IDSs are reported to have high detection rate and fast processing speed \cite{18-NIDvDL}. They are gaining increased significance alongside with the surge of network traffic and heterogenous types of network attacks. The performance of learning based IDSs \cite{15TON-NIDvML4Trafic} \cite{16-NIDvRF} has significant dependence on network data features and the choice of relevant classifiers (e.g., support vector machine (SVM), and logistic regression (LR)) which are obtained in a data-driven manner. In this case, the defects, such as feature redundance and data imbalance, of the training set become the bottleneck of developing accurate IDSs \cite{18-NIDvDL} \cite{16TOC-NIDvFeaSelect}.

In the past decade, deep learning (DL) models, e.g., deep neural networks (DNNs), have revolutionized classical ML on supervised classification tasks \cite{15-VGGNet} \cite{16-ResNet}. In the community of network security, DL based IDSs \cite{18-NIDvDL} \cite{16-NIDvDL4SDN} show advanced NID performance because hierarchically structured DNNs extract representative features of network data. Unfortunately, the weights of DNNs need to be optimized on a large dataset. Therefore, DL based IDSs cannot accurately detect network intrusions if the related training data are insufficient and imbalanced \cite{18-NIDvDL}.

The challenges encountered in learning based IDSs can be formulated as \emph{data scarcity} and \emph{data imbalance}. One highly potential solution of those problems is to increase the number of related data samples in the training set. However, labeling large datasets is expensive, time-consuming, and sometimes impossible due to emerging and fast evolving intrusion attacks. In addition, adequate records of different types of intrusions might be unavailable, which makes those problems particularly severe. Thus, developing a \emph{data augmentation} (DA) enhanced NID framework is crucial for network security.

To tackle the aforementioned challenges, this paper presents a novel NID framework that is able to utilize supervised classification models for identifying normal network requests and high-frequency cyber attacks. More importantly, we propose a DA module that incorporates deep adversarial learning and statistical learning techniques, which allows the NID framework to detect network intrusions in small sample scenarios. Experimental results on KDD Cup 99 dataset \cite{Data-KDDCup99} show that it can effectively identify intrusions, in particular, the emerging ones when limited training data are provided.

The main contributions of this paper are summarized as follows:

\begin{itemize}
  \item We propose a general learning based NID framework focusing on detecting network intrusion in small samples. By exploiting DA and advanced classification methods, it is capable of identifying a variety of network intrusions, especially emerging ones.
  \item We develop a novel DA module that addresses the scarcity and the imbalance of training set via a data-driven manner. It involves a probabilistic generative model for estimating network data feature distributions and generating synthesised data using Monte Carlo methods. Furthermore, we pre-train and fine-tune deep generative neural networks in adversarial learning scheme for augmenting synthesised intrusion data with high quality.
  \item Extensive experiments on classifying small sample intrusions and normal network requests are performed. Compared with the existing learning based IDSs, the DA enhanced NID framework achieves the better or comparable accuracy, precision, recall and F1-score.
\end{itemize}

The remainder of this paper is organized as follows: Section \ref{Section-RelatedWork} introduces related works and motivations. Section \ref{Section-OurWork} presents the proposed NID framework and learning based DA module. Section \ref{Section-Experiment} reports the evaluation, analysis, and comparison of experimental results. Section \ref{Section-Discussion} provides further discussions of our work. Section \ref{Section-Conclusion} concludes this paper.

\section{Related Work}
\label{Section-RelatedWork}

In this section, we first introduce recent progresses in learning based IDSs. We then present the preliminaries of DA methods and the motivations of this paper.

\subsection{Learning based Network Intrusion Detection}

Existing learning based IDSs utilize ML and DL models for distinguishing different types of network data. In the category of recent ML based IDSs, Zhang et al. \cite{15TON-NIDvML4Trafic} combined unsupervised clustering and supervised learning for robust network traffic classification. Ashfaq et al. \cite{17IS-NIDvFuzzy4BinClass} proposed semi-supervised fuzzy method for intrusion detection. Those methods establish NID models by learning knowledge from a large amount of unlabeled network data. Howbeit, the performance of their methods in detecting intrusions with small sample sizes remains unknown if abundant data are unavailable. Apart from those methods, supervised classification models such as LR, SVM, and random forest (RF) have been extensively applied for improving modern IDSs \cite{16TOC-NIDvFeaSelect} \cite{17ICBD-NIDvWorldEmbed} \cite{16-NIDvRF}. To deal with the feature redundancy problem in training supervised classifiers, Ambusaidi et al. \cite{16TOC-NIDvFeaSelect} introduced mutual information based algorithm to select network features. Zhou et al. \cite{17ICBD-NIDvWorldEmbed} adopted word embedding for extracting meaningful features of network data. In order to further alleviate the over-fitting problem, RF \cite{16-NIDvRF} and tree algorithms \cite{17NCA-NIDvTree} have been employed to ensemble sub-classification models for robust NID. However, the data scarcity and data imbalance remain unsolved because the feature selection/extraction and classifier aggregation do not increase the number of intrusion samples among different categories in the training set.

In DL based IDSs, Tang et al. \cite{16-NIDvDL4SDN} applied three-layer DNN for extracting multilevel features and classifying flow-based network intrusions. Shone et al. \cite{18-NIDvDL} proposed a more complex DNN called nonsymmetric deep auto-encoder (NDAE). They stacked two NDAEs for unsupervised feature learning of network data, followed by a RF for classification. Those DNN based IDSs achieve promising NID performance in terms of different evaluation metrics (e.g., accuracy, precision and recall). Considering the requirement of large quantities of training data for optimizing the layer weights of DNNs, the weakness of those IDSs in detecting low-frequency intrusions (e.g., the \verb"guesspasswd" and the \verb"bufferoverflow" attack reported in \cite{18-NIDvDL}) is non-negligible. Thus, enriching small sample intrusions in the training set is essential in developing learning based IDSs.

\subsection{Data Augmentation Methods}
\label{Section-DAmethods}

\subsubsection{Probabilistic Generative Models}

In Bayesian statistical learning, probabilistic generative models are extensively used to approximate unknown distributions of target data \cite{MCMC4ML}. Therein Markov chain Monte Carlo (MCMC) methods are used to sample their parameters from observed data.

The Metropolis-Hastings (MH) algorithm \cite{MH-M} is the foundation of MCMC, which generates candidate data from a proposal distribution based on an acceptance probability. An improved version to MH algorithm is Gibbs sampling \cite{GibbsSampling} \cite{ML-Prob-Gibbs} which uses a full conditional distribution for designing the proposal distribution. Closely related to Gibbs sampling, expectation maximisation (EM) algorithm \cite{MCMC-EM} \cite{MCMC4ML} alternates between an expectation step and a maximization step for estimating distribution parameters. Nonetheless, those algorithms cannot be directly applied to generate intrusions due to the difficulty of modeling complex features (e.g., real/Boolean values) with large divergence.

\subsubsection{Generative Adversarial Networks}

Generative adversarial networks (GANs) \cite{GANs2014} have been increasingly employed in generating realistic objects in computer vision \cite{GANs2D} \cite{GANs3D}. GANs contain two differently structured DNNs: the \emph{discriminative net} and the \emph{generative net} denoted as $D$ and $G$, respectively. During the stage of training, $G$ attempts to produce forged data based on input priority. $D$ takes in both forged data and real ones, and learns to distinguish the counterfeit from the truth. This finally results in a powerful data generator $G$.

Note that both $D$ and $G$ are constructed of DNNs, therefore, training GANs on a small collection of intrusion samples is a challenge. Even if sufficient network intrusions are available, GANs are reported to have remarkable training difficulties \cite{W-GAN} \cite{GANsTrain}. For instance, $G$ gets worse while $D$ gets better due to the significant difference of convergence speed, which might lead to poor outputs and a deficient generator.

\begin{figure*}[t]
  \centering
  \includegraphics[scale=0.57]{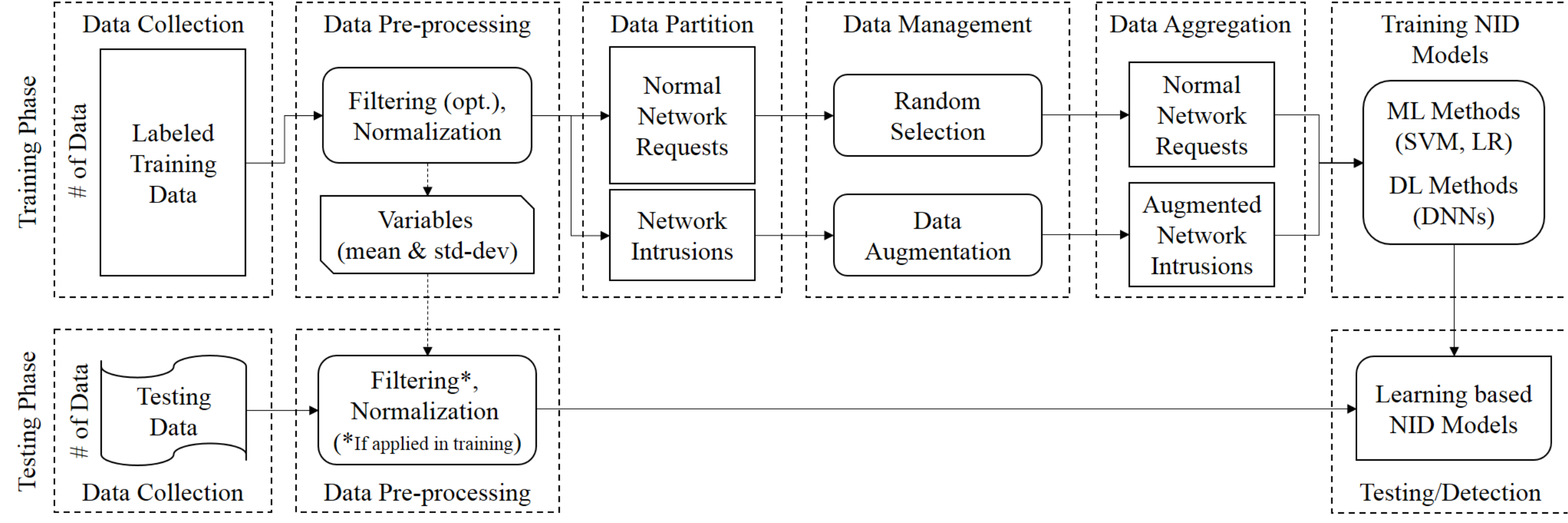}
  \caption{The proposed NID framework for detecting network intrusions in small sample sizes.}
  \label{Fig-NIDFramework}
\end{figure*}

\subsection{Motivations}

Despite promising progresses in recent research on learning based IDSs, challenges such as data scarcity and data imbalance in training (semi-)supervised classifiers are still two fundamental issues to the community. Those challenges become vital with the increased emerging attacks that often have limited known data samples. Data augmentation methods provide a potential approach to tackle these problems, which could enrich network data in the training set if designed properly. Findings obtained from the existing literatures show that directly applying DA algorithms might bring undesired effects \cite{MCMC4ML} \cite{GANsTrain}. Hence, a novel learning based DA module associated with the general NID framework presented in this paper will contribute to the design of advanced high performance IDSs.

\begin{figure}[t]
  \centering
  \includegraphics[scale=0.29]{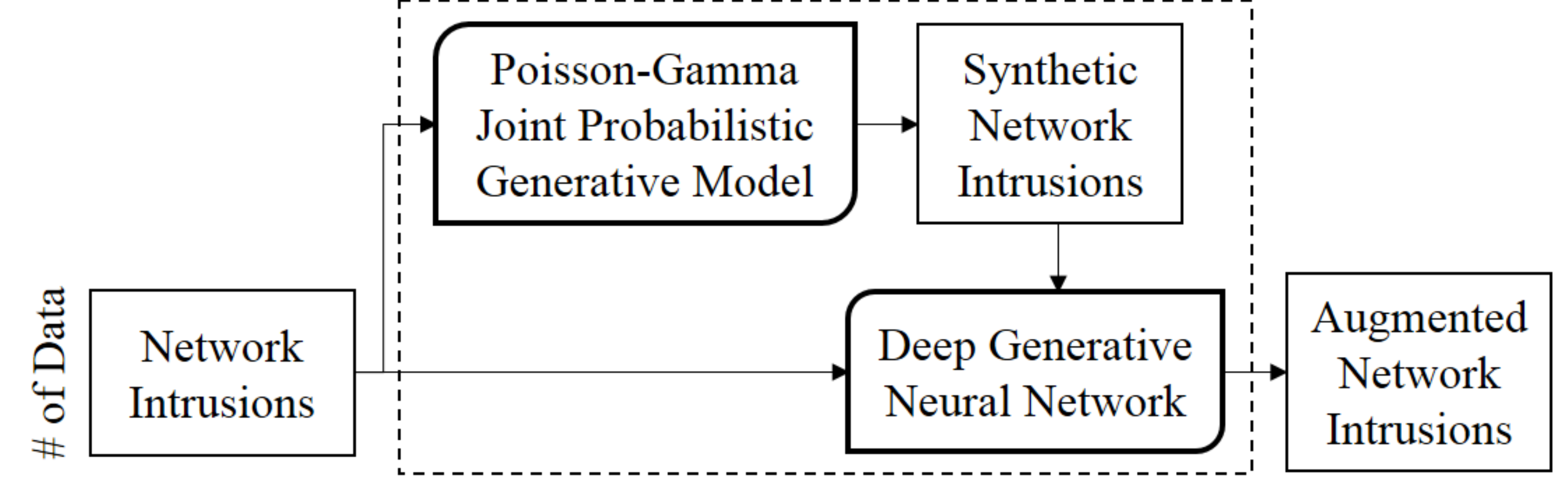}
  \caption{The structure of DA module for augmenting network intrusions.}
  \label{Fig-DAModule}
\end{figure}

\section{Network Intrusion Detection with Data Augmentation}
\label{Section-OurWork}

In this section, we first introduce the pipeline of the proposed NID framework and then present the formulation of the learning based DA module for addressing data scarcity and imbalance problems. Finally, we provide the detailed optimization of DA module through adversarial learning.

\subsection{The NID Framework}

Figure \ref{Fig-NIDFramework} depicts the proposed NID framework that involves the \emph{training phase} for training supervised classification models assisted with learning based DA module. Those classifiers are then adopted in the \emph{testing phase} for detecting network intrusions from normal network requests.

In the \emph{training phase}, continuous features of network data are normalized and the related statistical variables (i.e., mean and standard deviation) are recorded for processing testing data. We make discrete symbolic features remain unchanged. Other pre-processing techniques (e.g., data filtering and feature selection) can be alternatively applied. Normalized data are then partitioned, in which a proportion of normal network data are randomly selected and the imbalanced network intrusion samples are augmented by DA module. Finally, those data are aggregated as a balanced dataset for training NID models using supervised learning methods.

In the \emph{testing phase}, input network data are pre-processed as training data and fed into NID models for classification. Because those models are trained on balanced dataset using advanced learning methods, the proposed framework is competent to accurately identify network intrusions, especially emerging ones.


\subsection{The Data Augmentation Module}

Figure \ref{Fig-DAModule} shows the schematic diagram of the proposed DA module. Given a small number of network intrusion data samples, the Poisson-Gamma joint probabilistic model (PGM) is first derived for efficiently generating synthesised intrusion data. Deep generative neural networks (DGNNs) then take both real and synthesised data to train layer weights through adversarial learning. Finally, DGNNs output network intrusion data with augmented quality.

In DA module, PGM plays an essential role in initializing DGNNs to fit in with the general distribution of related intrusions. This alleviates the convergence problem of training DGNNs on limited network data. In turn, DGNNs compensate for the boundedness of PGM on simulating network features that exhibit large divergence.

\subsection{The Poisson-Gamma Joint Probabilistic Model}

In this subsection, we present the Poisson-Gamma joint probabilistic generative model for modeling feature distributions of network data. At the same time, we theoretically analyze its feasibility for simulating complex network features. Finally, we exploit MCMC based Gibbs sampler for efficiently generating synthesised intrusion data.

Let $\bm{x} \in \mathbb{R}^{K}$ be the feature vector of one network intrusion sample, where $K$ denotes the dimension of feature space. Assuming that intrusions of the same category come from the joint probabilistic distribution defined as follows:

\begin{equation}\label{Eq-PGM-MixFeaDistrib}
  \ p_{\mathcal{M}}(\bm{x}) = p(\bm{x}|\bm{\lambda}),
\end{equation}

\noindent where $\bm{\lambda} = \{\bm{\lambda}_1, \dots, \bm{\lambda}_K\}$, and $\bm{\lambda}_i \in \mathbb{R}^{Q}$ denotes the distribution parameters of $i$-th feature dimension of $\bm{x}$. Note that $p_{\mathcal{M}}(\bm{x})$ contains $K$ components, which means each feature of $\bm{x}$ is jointly modeled by one distribution.

In the existing NID benchmarks \cite{Data-KDDCup99} \cite{Data-NSLKDD}, \emph{continuous} digital features of network data usually represent the volume of requests or the time of connections. Consequently, Poisson distribution \cite{StochasticProcessBook} is employed to approximate the accumulation of those network events:

\begin{equation}\label{Eq-PGM-Distrib-Poisson}
  \text{Poisson}(\bm{x}; \bm{\lambda}) = e^{-\bm{\lambda}} \frac{\bm{\lambda}^{\bm{x}}}{\bm{x}!},
\end{equation}

\noindent where $\bm{\lambda}$ indicates the average intensity, i.e., the statistical average, of $\bm{x}$. Since Poisson distribution contains one parameter, we have $\bm{\lambda}_i \in \mathbb{R}^{Q}$ and $Q = 1$.

To render an effective Markov chain for the subsequent data simulation, $\bm{\lambda}$ is approximated by Gamma distribution:

\begin{equation}\label{Eq-PGM-Distrib-Gamma}
  \text{Gamma}(\bm{\lambda}; \bm{a}, \bm{b}) = \frac{1}{\Gamma(\bm{a})} \bm{b}^{\bm{a}} \lambda^{\bm{a}-\bm{I}} \text{exp}(-\bm{b}\bm{\lambda}),
\end{equation}

\noindent where $\bm{a}, \bm{b} \in \mathbb{R}^{K}$ are shape and scale parameters. Given a collection of $M$ intrusion samples $\bm{X} \in \mathbb{R}^{M \times K}$, we have $\bm{a} = [E(\bm{X})]^{2}/Var(\bm{X})$ and $\bm{b} = E(\bm{X})/Var(\bm{X})$, where $E(\bm{X})$ and $Var(\bm{X})$ denote the expectation and variance of $\bm{X}$. Because of the conjugated relationship between Poisson and Gamma distribution \cite{MCMCinPractice}, the convergence property of PGM is theoretically guaranteed for synthesising network data.

\begin{prop}
 The Poisson-Gamma joint probabilistic model can estimate the distributions of both continuous and discrete digital features of network data.
\end{prop}

\begin{prof}
 The Poisson-Gamma joint distribution can approximately estimate \emph{continuous} digital features distributions since the accumulation of network events can be statistically formulated by Poisson process.

 For \emph{discrete} digital features (i.e., values equals to 0 or 1), the joint distribution can model those symbolic values by sampling the Poisson parameter $\bm{\lambda}_{i}$ to be 0 or 1 from Gamma distribution, respectively. \qed
\end{prof}

Providing intrusion samples $\bm{X} = \{\bm{x}^{(i)}\}_{i=1}^{M}$ of the same category, the goal of synthesising intrusions then becomes to estimate $\bm{\lambda}$ given $\bm{X}$. According to Bayes' theorem \cite{PatternClassification}, this can be achieved by maximizing the compact posterior:

\begin{equation}\label{Eq-PGM-Post}
  \ p(\bm{\lambda}|\bm{X}) \propto \ p(\bm{X}|\bm{\lambda})p(\bm{\lambda}),
\end{equation}

\noindent where the first and the second terms on the right side are the density functions of Poisson and Gamma distributions, respectively.

\begin{prop}
 If network intrusion samples are assumed to obey Poisson distribution and their Poisson parameter $\bm{\lambda}$ is formulated by Gamma distribution, then the posterior $p(\bm{\lambda}|\bm{X})$ obeys Gamma distribution.
\end{prop}

\begin{prof}
 In Eq. (\ref{Eq-PGM-Post}), each term stands for a probability and hence has a nonnegative value. Applying natural logarithm on both sides in Eq. (\ref{Eq-PGM-Post}), we have:

 \begin{equation}\label{Eq-PGM-PostLog}
   \text{log}p(\bm{\lambda}|\bm{X})
   \propto \text{log}(\prod_{i=1}^{M}p(\bm{x}^{(i)}|\bm{\lambda})) + \text{log}p(\bm{\lambda}),
 \end{equation}

 \noindent where $i$ denotes the serial number of known intrusion samples.

 Substituting the exponential form of Poisson and Gamma density functions into Eq. (\ref{Eq-PGM-PostLog}), we obtain:

 \begin{equation}\label{Eq-PGM-PostLogExpand}
 \begin{aligned}
    & \text{log}(\prod_{i=1}^{M}p(\bm{x}^{(i)}|\bm{\lambda})) + \text{log}p(\bm{\lambda}) \\
  = & \ \sum_{i=1}^{M}( \bm{x}^{(i)} \text{log}\bm{\lambda} - \bm{\lambda} - \text{log}(\bm{x}^{(i)}!) ) \\
  + & \ (\bm{a}-\bm{I})\text{log}\bm{\lambda} - \bm{b}\bm{\lambda} -\text{log}\Gamma(\bm{a}) + \bm{a}\text{log}\bm{b},
 \end{aligned}
 \end{equation}

 \noindent where $\bm{x}!$ denotes the element-wise factorial of $\bm{x}$ and $\bm{I} \in \mathbb{R}^{K}$ a unit vector.

 Extracting the terms with respect to $\bm{\lambda}$ in Eq. (\ref{Eq-PGM-PostLogExpand}), we derive the representation of the posterior as follows:

 \begin{equation}\label{Eq-PGM-PostFinal}
 \begin{aligned}
    & \ \text{log}p(\bm{\lambda}|\bm{X}) \\
  = & \ C[\sum_{i=1}^{M}(\bm{x}^{(i)} \text{log}\bm{\lambda} - \bm{\lambda}) + (\bm{a}-\bm{I})\text{log}\bm{\lambda} - \bm{b}\bm{\lambda}] \\
  = & \ C[(\bm{a} + \sum_{i=1}^{M}\bm{x}^{(i)} - \bm{I})\text{log}\bm{\lambda} - (M\bm{I} + \bm{b})\bm{\lambda}] \\
  = & \ C[\text{log}\ \text{Gamma}(\bm{a} + \sum_{i=1}^{M}\bm{x}^{(i)}, M\bm{I} + \bm{b})],
 \end{aligned}
 \end{equation}

 \noindent where $C$ is a negligible constant \cite{MCMCinPractice}. All multiplications above are operated in an element-wise manner. Therefore, the posterior $p(\bm{\lambda}|\bm{X})$ obeys Gamma distribution.

\qed
\end{prof}

Given a collection of network intrusion samples $\{\bm{x}^{(i)}\}_{i=1}^{M}$, we utilize Eq. (\ref{Eq-PGM-PostFinal}) to approximate the Poisson parameter $\bm{\lambda}$ and further synthesise intrusion data by Poisson using Gibbs sampler. Algorithm \ref{Algo-PGM} shows the pseudo-code of PGM for producing synthesised intrusion data.

\begin{algorithm}[ht]
  \caption{The Poisson-Gamma joint probabilistic model}
  \label{Algo-PGM}
  \begin{algorithmic}[1]
    \Require
      A set of real intrusion samples $\{\bm{x}^{(i)}\}_{i=1}^{M}$,
      The number of synthesised intrusion data $N$,
      The cut off threshold of Gibbs sampler $T$.
    \Ensure
      A set of synthesised intrusion data $\{\bm{y}^{(j)}\}_{j=1}^{N}$.
    \State Initialize $n = 1$;
    \For{$t$ = $1,\dots, N+T$}
      \State Sample $\bm{\lambda}^{(t)} \sim \text{Gamma}(; \{\bm{x}^{(i)}\}_{i=1}^{M},\bm{a},\bm{b})$ in Eq. (\ref{Eq-PGM-PostFinal});
      \If{$t > T$}
        \State Sample $\bm{y}^{(n)} \sim \text{Poisson}(; \bm{\lambda}^{(t)})$ in Eq. (\ref{Eq-PGM-Distrib-Poisson});
        \State $n = n + 1$;
      \EndIf
    \EndFor
  \end{algorithmic}
\end{algorithm}

\begin{figure}[t]
  \centering
  \subfigure[The pre-training process (forward-propagation).]{
  \includegraphics[scale=0.45]{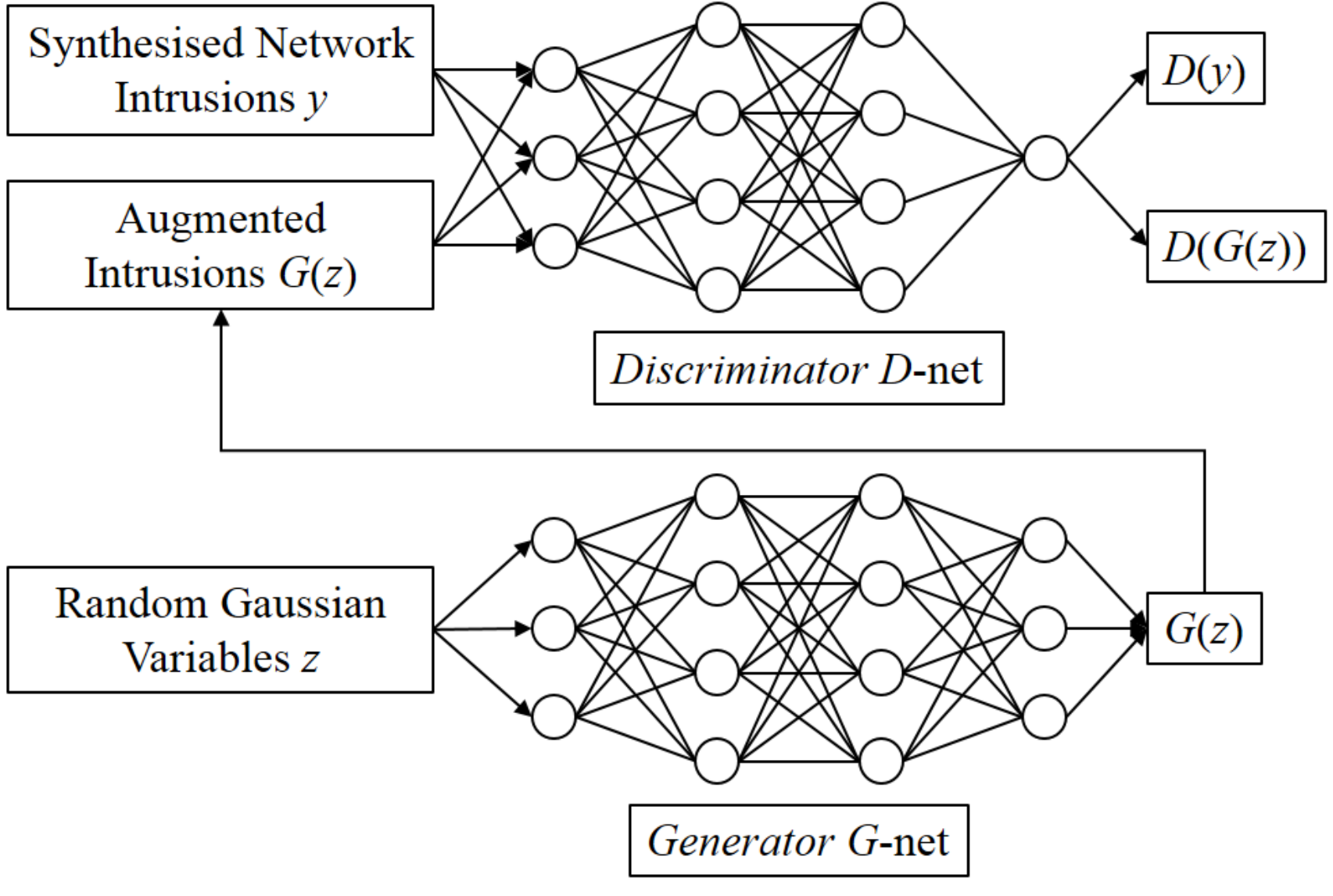}}
  \subfigure[The fine-tuning process (forward-propagation).]{
  \includegraphics[scale=0.45]{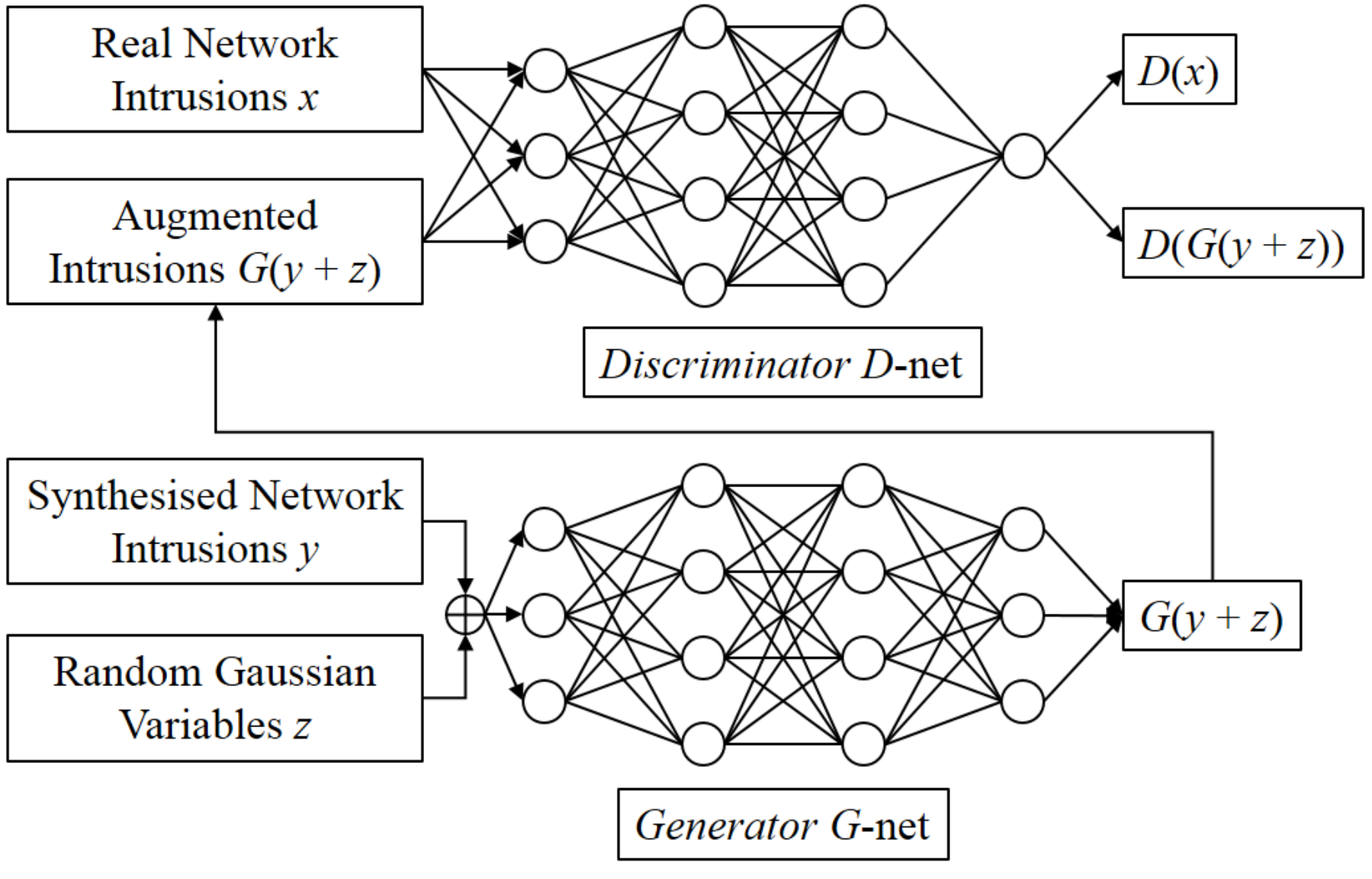}}
  \caption{The schematic depiction of two-fold adversarial mechanism for training DGNNs.}
  \label{Fig-DGNNs}
\end{figure}

\subsection{The Deep Generative Neural Networks}

In this subsection, we present the formulation of DGNNs that augments synthesised intrusion data with enhanced quality. Afterwards, we propose a two-fold mechanism for training DGNNs via adversarial learning strategies.

The proposed DGNNs have two components: the \emph{Discriminator} ($D$-net) and the \emph{Generator} ($G$-net). In adversarial training process, $G$-net generates augmented intrusion data by learning their real feature distribution, while $D$-net acts as an indicator trying to reject augmented data from real intrusion samples. By fully exploiting the capability of hierarchical feature learning, $D$-net and $G$-net are implemented by DNNs. The input size of $D$-net and $G$-net is $K$ which equals to the dimension of network data features. In the hidden layers, neural nodes are stacked for abstracting latent representation of input data. The output of $D$-net is a scalar which returns the possibility whether its input comes from real samples. In contrast, the output of $G$-net is a $K$-dimensional vector representing the augmented network intrusion.

Provided with target network intrusion samples, the goal of training DGNNs is to obtain a potent $G$-net that recovers their feature distributions and further generates intrusion data with similar characters. As discussed in Section \ref{Section-DAmethods}, training DGNNs with limited samples is easy to over-fit because both $D$-net and $G$-net are structured in DNNs. Therefore, we present a two-fold adversarial learning mechanism which adopts both real and synthesised network data for optimizing their weights.

\subsubsection{Pre-training}

As shown in Figure \ref{Fig-DGNNs}(a), synthesised intrusion data $\bm{Y} = \{\bm{y}^{(j)}\}_{j=1}^{N}$ are treated as sub-optimal targets and $D$-net is trained to distinguish augmented ones (i.e., $G(\bm{Z})$) from them. Variables $\bm{Z} = \{\bm{z}^{(j)}\}_{j=1}^{N}$ and $\bm{z} \in \mathbb{R}^{K}$ are the prior of $G$-net for learning the mapping $G(\bm{Z}) \to \bm{Y}$. To adjust to different and unexpected situations, $\bm{z}$ is sampled from Gaussian distribution $\mathcal{N}(0, 1)$. This assumes $G$-net has no specific prior knowledge on targets. The objective of pre-training is then formed as the following adversarial game:

\begin{equation}\label{Eq-DGNNs-pretrain}
\begin{aligned}
    & \ \text{min}_G \ \text{max}_D \ V(D, G) \\
  = & \ \bbbe_{\bm{y} \sim p_{\text{fake}}}[\text{log}D(\bm{y})] + \bbbe_{\bm{z} \sim p_{\text{noise}}}[\text{log}(1-D(G(\bm{z})))],
\end{aligned}
\end{equation}

\noindent where $p_{\text{fake}}$ and $p_{\text{noise}}$ denote feature distributions of $\bm{Y}$ and $\bm{Z}$, respectively. Given sufficient synthesised data, DGNNs are pre-trained to fit in with the general distribution of related network intrusions \cite{GANsTrain}. In this case, the quality of augmented samples is bound to be mediocre as $G$-net cannot directly access the feature information of real intrusion samples.

\subsubsection{Fine-tuning}

During the adversarial training demonstrated in Figure \ref{Fig-DGNNs}(b), $D$-net takes in both real intrusion samples $\bm{X} = \{\bm{x}^{(i)}\}_{i=1}^{M}$ and augmented ones from $G$-net. Meanwhile, $G$-net is fed by synthesised data added with Gaussian variables (i.e., $\bm{y}+\bm{z}$). In this scenario, DGNNs are fine-tuned according to the following objective:

\begin{equation}\label{Eq-DGNNs-finetune}
\begin{aligned}
    & \ \text{min}_G \ \text{max}_D \ V(D, G) \\
  = & \ \bbbe_{\bm{x} \sim p_{\text{real}}}[\text{log}D(\bm{x})] \\
  + & \ \bbbe_{\bm{y} \sim p_{\text{fake}}, \bm{z} \sim p_{\text{noise}}}[\text{log}(1-D(G(\bm{y}+\bm{z})))],
\end{aligned}
\end{equation}

\noindent where $p_{\text{real}}$ denotes the real intrusion distribution to recover. Note that $G(\bm{y}+\bm{z})$ rather than $G(\bm{y})$ or $G(\bm{z})$ is chosen in fine-tuning. This avoids $G$-net learning monotonous distribution from the same data in $\bm{Y} = \{\bm{y}^{(j)}\}_{j=1}^{N}$.

Since DGNNs have been pre-trained on synthesised intrusion data, the \emph{Generator} competes with the \emph{Discriminator} on a comparative level in fine-tuning stage. Thus, the two-fold adversarial training mechanism allows DGNNs to learn real intrusion distribution in a progressive manner. This precludes the remarkable convergence difference of DGNNs, and prevents the \emph{Generator} generating poor outputs.

\begin{algorithm}[ht]
  \caption{Pre-train and fine-tune DGNNs in adversarial manner with stochastic gradient optimization for augmenting synthesised network intrusion data}
  \label{Algo-DGN}
  \begin{algorithmic}[1]
    \Require
       Real intrusion samples $\{\bm{x}_i\}_{i=1}^{M}$,
       Synthesised intrusion data $\{\bm{y}_j\}_{j=1}^{N}$,
       Gaussian random variables $\{\bm{z}_j\}_{j=1}^{N}$,
       Mini-batch size $L$,
       Maximum iteration $T_{\text{pre-train}}$ and $T_{\text{fine-tune}}$.
    \Ensure
       Optimized weights $\theta_d$ and $\theta_g$ of DGNNs.
    \For{$t = 1,2,\dots,T_{\text{pre-train}}+T_{\text{fine-tune}}$}
      \If{$t \leq T_{\text{pre-train}}$}
        \For{$k_d$ steps}
          \State Select $L$ samples from $\{\bm{y}_j\}_{j=1}^{N}$;
          \State Select $L$ samples from $\{\bm{z}_j\}_{j=1}^{N}$;
          \State Compute gradient $\nabla_{\theta_d}$ as Eq. (\ref{Eq-DGNNs-Grad-D-pretrain});
          \State Update $\theta_d$ by Adam.
        \EndFor
        \For{$k_g$ steps}
          \State Select $L$ samples from $\{\bm{z}_j\}_{j=1}^{N}$;
          \State Compute gradient $\nabla_{\theta_g}$ as Eq. (\ref{Eq-DGNNs-Grad-G-pretrain});
          \State Update $\theta_g$ by Adam.
        \EndFor
      \Else
        \For{$k_d$ steps}
          \State Select $L$ real intrusions from $\{\bm{x}_i\}_{i=1}^{M}$;
          \State Select $L$ samples from $\{\bm{y}_j\}_{j=1}^{N}$;
          \State Select $L$ samples from $\{\bm{z}_j\}_{j=1}^{N}$;
          \State Compute gradient $\nabla_{\theta_d}$ as Eq. (\ref{Eq-DGNNs-Grad-D-finetune});
          \State Update $\theta_d$ by Adam.
        \EndFor
        \For{$k_g$ steps}
          \State  Select $L$ samples from $\{\bm{y}_j\}_{j=1}^{N}$;
          \State  Select $L$ samples from $\{\bm{z}_j\}_{j=1}^{N}$;
          \State  Compute gradient $\nabla_{\theta_g}$ as Eq. (\ref{Eq-DGNNs-Grad-G-finetune});
          \State  Update $\theta_g$ by Adam.
        \EndFor
      \EndIf
    \EndFor
  \end{algorithmic}
\end{algorithm}

\subsection{Optimization}

In this section, we present the optimization details of pre-training and fine-tuning DGNNs.

Let $\theta_d$ and $\theta_g$ denote the weights of $D$-net and $G$-net. The aim of training DGNNs becomes to optimize the minmax objectives in Eq. (\ref{Eq-DGNNs-pretrain}) and Eq. (\ref{Eq-DGNNs-finetune}) with respect to $\theta_d$ and $\theta_g$. Considering that DGNNs are formed in DNNs, they are trained by back-propagation with stochastic gradient algorithm.

In the pre-training stage, the gradients of $V$ with respect to $\theta_d$ and $\theta_g$ on a batch of $L$ intrusion samples are:

\begin{equation}\label{Eq-DGNNs-Grad-D-pretrain}
    \frac{\partial V}{\partial \theta_d}
    = \nabla_{\theta_d} \frac{1}{L} \sum_{l=1}^{L} [\text{log}D(\bm{y}^{(l)}) + \text{log}(1 - D(G(\bm{z}^{(l)})))],
\end{equation}

\begin{equation}\label{Eq-DGNNs-Grad-G-pretrain}
  \frac{\partial V}{\partial \theta_g}
  = \nabla_{\theta_g} \frac{1}{L} \sum_{l=1}^{L} [\text{log}(1 - D(G(\bm{z}^{(l)})))].
\end{equation}

In the fine-tuning stage, the gradients of $V$ with respect to $\theta_d$ and $\theta_g$ on a batch of $L$ intrusion samples are:

\begin{equation}\label{Eq-DGNNs-Grad-D-finetune}
    \frac{\partial V}{\partial \theta_d}
    = \nabla_{\theta_d} \frac{1}{L} \sum_{l=1}^{L} [\text{log}D(\bm{x}^{(l)}) + \text{log}(1 - D(G(\bm{y}^{(l)}+\bm{z}^{(l)})))],
\end{equation}

\begin{equation}\label{Eq-DGNNs-Grad-G-finetune}
  \frac{\partial V}{\partial \theta_g}
  = \nabla_{\theta_g} \frac{1}{L} \sum_{l=1}^{L} [\text{log}(1 - D(G(\bm{y}^{(l)}+\bm{z}^{(l)})))].
\end{equation}

\noindent After obtaining gradients ${\partial V}/{\partial \theta_d}$ and ${\partial V}/{\partial \theta_g}$, layer weights $\theta_d$ and $\theta_g$ are optimized by stochastic gradient descent.

Algorithm \ref{Algo-DGN} shows the pseudo-code of training DGNNs. During the evolution of adversarial learning, we alternate between $k_d = 8$ steps of optimizing $D$-net and $k_g = 2$ steps of optimizing $G$-net. In each inner loop, the gradient descent method Adam \cite{ADAM} is employed for updating $\theta_d$ and $\theta_g$.

\section{Experiments}
\label{Section-Experiment}

In this section, we conduct comprehensive experimental validations of the proposed DA enhanced NID framework on KDD Cup 99 dataset \cite{Data-KDDCup99}. Its performance on detecting the network intrusions in small samples (e.g., emerging attacks) is compared with learning based IDSs, in which classical LR, SVM, and advanced DNN are employed for comparison.

\subsection{Network Data}

KDD Cup 99 dataset \cite{Data-KDDCup99} is a benchmark dataset widely used in NID study. As outlined in Table \ref{Tab-Data-KDDCup99}, it is composed of two training sets at different scales and one testing set. Network data are categorized into normal requests (NORMAL) and four major intrusions: denial-of-service (DOS), surveillance and other probing (PROBE), unauthorized access from a remote machine (R2L) and unauthorized access to root user (U2R). Given that 100\% training set merely contains additional records of normal requests and high-frequency intrusions, the 10\% training set is used in all experiments.

\begin{table}[t]
\centering
\setlength{\abovecaptionskip}{0pt}
\setlength{\belowcaptionskip}{0pt}
  \caption{KDD Cup 99 Dataset (No. of samples in each category)}
  \label{Tab-Data-KDDCup99}
  \renewcommand\arraystretch{1.1}
  \centering
  \begin{tabular}{*{4}{l c c c}}
  \hline
  \hline
  \specialrule{0em}{2pt}{2pt}
    \textbf{Category}    &  \textbf{Training (100\%)}  &  \textbf{Training (10\%)}  &  \textbf{Testing}  \\
  \specialrule{0em}{2pt}{2pt}
  \hline
  \specialrule{0em}{1pt}{1pt}
    \textbf{NORMAL}      &  972781            &  97278            &  60593    \\
    \textbf{DOS}         &  3883370           &  391458           &  229853   \\
    \textbf{PROBE}       &  41102             &  4107             &  4166     \\
    \textbf{U2R}         &  52                &  52               &  228      \\
    \textbf{R2L}         &  1126              &  1126             &  16189    \\
  \specialrule{0em}{1pt}{1pt}
  \hline
  \hline
  \end{tabular}
\end{table}

\begin{table}[t]
\centering
\setlength{\abovecaptionskip}{0pt}
\setlength{\belowcaptionskip}{0pt}
  \caption{No. of training samples in the subcategories that have severe scarcity and imbalance problems in the KDD Cup 99 Dataset}
  \label{Tab-Data-8Intrusions}
  \renewcommand\arraystretch{1.1}
  \centering
    \begin{tabular}{*{6}{l l c c}}
    \hline
    \hline
    \specialrule{0em}{2pt}{2pt}
    \multicolumn{1}{l}{{\textbf{Category}}} & \multicolumn{1}{l}{{\textbf{Intrusion Type}}} & \multicolumn{1}{c}{\textbf{Training (10\%)}} & \multicolumn{1}{c}{\textbf{Testing}}  \\
    \specialrule{0em}{2pt}{2pt}
    \hline
    \specialrule{0em}{1pt}{1pt}
    \multirow{3}*{\textbf{DOS}}
    & 'apache2'        & 0   & 794    \\
    & 'mailbomb'       & 0   & 5000   \\
    & 'processtable'   & 0   & 759    \\
    \specialrule{0em}{1pt}{1pt}
    \hline
    \specialrule{0em}{1pt}{1pt}
    \multirow{2}*{\textbf{PROBE}}
    & 'mscan'          & 0   & 1053   \\
    & 'saint'          & 0   & 736    \\
    \specialrule{0em}{1pt}{1pt}
    \hline
    \specialrule{0em}{1pt}{1pt}
    \multirow{3}*{\textbf{R2L}}
    & 'guesspasswd'    & 53  & 4367   \\
    & 'snmpgetattack'  & 0   & 7741   \\
    & 'snmpguess'      & 0   & 2406   \\
    \specialrule{0em}{1pt}{1pt}
    \hline
    \hline
    \end{tabular}
\end{table}

Each record of KDD Cup 99 dataset consists of 38 digital features and 4 character features. Assuming that the digital features provide adequate information for identification, characters are not used for experimental validation.

Considering the task of NID in small sample sizes, network intrusions are expected to satisfy the prerequisites that they have limited records in the training set, while plentiful records exist in the testing set. In this case, 8 types of intrusions are selected (as shown in Table \ref{Tab-Data-8Intrusions}): \verb"apache2", \verb"mailbomb" and \verb"processtable" of DOS attack, \verb"mscan" and \verb"saint" of PROBE attack, \verb"guesspasswd", \verb"snmpgetattack" and \verb"snmpguess" of R2L attack. In experiments, if no training data in one intrusion type is available, $M$ related testing intrusion samples will be selected to complement the training set before data augmentation. Those samples are then rejected for testing. Despite training DNNs requires a large amount of labelled data, $M$ is set to 50 to meet the above prerequisites.

\subsection{Evaluation Metrics}

The metrics that are used to measure NID results of IDSs are listed below:

\begin{itemize}
  \item True Positive (\emph{TP}): Network intrusions (or normal network requests) that are correctly detected.
  \item True Negative (\emph{TN}): Normal network requests (or network intrusions) that are correctly detected.
  \item False Positive (\emph{FP}): Normal requests that are mis-classified as intrusions.
  \item False Negative (\emph{FN}): Intrusions that are mis-classified as normal requests.
\end{itemize}

Then, the accuracy, precision, recall, and F1-score \cite{18-NIDvDL} \cite{16-NIDvDL4SDN} are computed to evaluate different IDSs, in which larger values represent the better detection performance on network data.

\subsection{Parameters}

The detailed parameter setup and tuning strategies are provided as follows:

\subsubsection{Threshold of Gibbs sampler}

The cut off threshold $T$ is set to be 500 to assure the convergence of PGM. Due to the efficiency of Gibbs sampling, $T$ can be a large value while the time consuming is still economic.

\subsubsection{Structure of DGNNs}

In $D$-net, hidden nodes are set to be 70, 50, 40, and 20. The ReLU activation function is used after each non-terminal layer, while the the sigmoid activation function is applied to the last layer for producing the decision probability. In $G$-net, the layer nodes are set to be 40, 30, and 20. Those three hidden layers are sequentially connected by ReLu and sigmoid functions. The last hidden layer is linearly mapped to the output layer. In this scenario, $G$-net has less hidden nodes than that of $D$-net, which leads to improved inference capacity of $D$-net and guarantees an effective $G$-net to be trained. In addition, dropout \cite{Dropout} is employed in all hidden layers to regularize training process and decrease over-fitting risks.

\subsubsection{Mini-batch size of DGNNs}

$L$ is constrained to be less than the minimum number of real and synthesised intrusion data (i.e., $L < \text{min}\{M, N\}$). In our experiments, $L$ is set to be 20\% $\sim$ 40\% of real intrusion samples (i.e., $L = 15$). This allows training DGNNs to be efficiently and effectively.

\subsubsection{Training iterations of DGNNs}

The maximum iterations in pre-training stage and fine-tuning stage are empirically set to be $T_{\text{pre-train}} = 3000$ and $T_{\text{fine-tune}} = 300$. Since known network intrusions are limited, less iterations in fine-tuning stage are suggested to avoid over-fitting problem.

\subsubsection{Other hyper-parameters of DGNNs}

The learning rates of $D$-net and $G$-net are both empirically set to be $2 \times 10^{-4}$ for the propose of adjusting weights of DNNs with respect to low ratio of loss gradients. This guarantees DNNs to be robustly trained and avoids missing optimum. Howbeit, it might result in low convergence speed. Thus, momentum \cite{Momentum} is adopted to accelerate training and prevent gradient oscillations. The parameters in momentum are defaultly set to be 0.9 and 0.99 when Adam optimizer is used \cite{ADAM}.

\subsection{Binary Classification based NID}

\begin{table*}[t]
\centering
\setlength{\abovecaptionskip}{0pt}
\setlength{\belowcaptionskip}{0pt}
  \caption{ML based 2-class NID with LR and SVM\protect\linebreak(Mean $\pm$ Std-Dev Percent)}
  \label{Tab-Res2C-LRSVM}
  \renewcommand\arraystretch{1.1}
  \centering
    \begin{tabular}{*{6}{p{2cm} p{2.0cm} c c c c}}
    \hline
    \hline
    \specialrule{0em}{2pt}{2pt}
    \multicolumn{1}{l}{{\textbf{Intrusion Type}}} & \multicolumn{1}{l}{\textbf{NID Model Name}} & \multicolumn{1}{c}{\textbf{Accuracy}} & \multicolumn{1}{c}{\textbf{Precision}} & \multicolumn{1}{c}{\textbf{Recall}} & \multicolumn{1}{c}{\textbf{F1-Score}} \\
    \specialrule{0em}{2pt}{2pt}
    \hline
    \specialrule{0em}{1pt}{1pt}
    \multirow{4}*{'apache2'}
    & NID-LR                     & 98.51 $\pm$ 0.42   & 0.00               & 0.00               & -                  \\
    & NID-\textbf{DA}-LR         & \textbf{99.53} $\pm$ \textbf{0.05}   & \textbf{77.29} $\pm$ \textbf{1.21}   & \textbf{90.70} $\pm$ \textbf{3.41}   & \textbf{83.45} $\pm$ \textbf{2.14}   \\
    \specialrule{0em}{1pt}{1pt}
    \cline{2-6}
    \specialrule{0em}{1pt}{1pt}
    & NID-SVM                    & 98.97 $\pm$ 0.02   & 55.73 $\pm$ 0.56   & 99.32 $\pm$ 0.38   & 71.39 $\pm$ 0.49   \\
    & NID-\textbf{DA}-SVM        & \textbf{99.94} $\pm$ \textbf{0.01}   & \textbf{95.61} $\pm$ \textbf{0.87}   & \textbf{99.70} $\pm$ \textbf{0.07}   & \textbf{97.61} $\pm$ \textbf{0.44}   \\
    \specialrule{0em}{1pt}{1pt}
    \hline
    \specialrule{0em}{1pt}{1pt}
    \multirow{4}*{'mailbomb'}
    & NID-LR                     & 91.93 $\pm$ 0.44   & 0.00               & 0.00               & -                  \\
    & NID-\textbf{DA}-LR         & \textbf{97.84} $\pm$ \textbf{0.41}   & \textbf{78.11} $\pm$ \textbf{3.37}   & \textbf{99.89} $\pm$ \textbf{0.14}   & \textbf{87.63} $\pm$ \textbf{2.08}   \\
    \specialrule{0em}{1pt}{1pt}
    \cline{2-6}
    \specialrule{0em}{1pt}{1pt}
    & NID-SVM                    & 93.04 $\pm$ 0.03   & 80.38 $\pm$ 1.58   & 11.53 $\pm$ 0.33   & 20.16 $\pm$ 0.50   \\
    & NID-\textbf{DA}-SVM        & \textbf{99.34} $\pm$ \textbf{0.34}   & \textbf{92.29} $\pm$ \textbf{3.65}   & \textbf{99.78} $\pm$ \textbf{0.16}   & \textbf{95.86} $\pm$ \textbf{2.05}   \\
    \specialrule{0em}{1pt}{1pt}
    \hline
    \specialrule{0em}{1pt}{1pt}
    \multirow{4}*{'processtable'}
    & NID-LR                     & 99.33 $\pm$ 0.05   & 64.98 $\pm$ 1.78   & 98.74 $\pm$ 2.03    & 78.37 $\pm$ 1.72   \\
    & NID-\textbf{DA}-LR         & \textbf{99.53} $\pm$ \textbf{0.04}   & \textbf{72.46} $\pm$ \textbf{1.81}   & \textbf{100.00} $\pm$ \textbf{0.00}   & \textbf{84.02} $\pm$ \textbf{1.21}   \\
    \specialrule{0em}{1pt}{1pt}
    \cline{2-6}
    \specialrule{0em}{1pt}{1pt}
    & NID-SVM                    & 99.87 $\pm$ 0.06   & 90.79 $\pm$ 3.86   & 99.60 $\pm$ 0.56    & 94.96 $\pm$ 2.14   \\
    & NID-\textbf{DA}-SVM        & \textbf{99.90} $\pm$ \textbf{0.04}   & \textbf{92.60} $\pm$ \textbf{2.17}   & \textbf{99.71} $\pm$ \textbf{0.15}    & \textbf{96.02} $\pm$ \textbf{1.22}   \\
    \specialrule{0em}{1pt}{1pt}
    \hline
    \specialrule{0em}{1pt}{1pt}
    \multirow{4}*{'mscan'}
    & NID-LR                     & 97.78 $\pm$ 0.11   & 42.58 $\pm$ 1.35   & 84.90 $\pm$ 1.48   & 56.71 $\pm$ 1.42   \\
    & NID-\textbf{DA}-LR         & \textbf{99.01} $\pm$ \textbf{0.06}   & \textbf{64.81} $\pm$ \textbf{1.48}   & \textbf{92.29} $\pm$ \textbf{0.46}   & \textbf{76.14} $\pm$ \textbf{1.14}   \\
    \specialrule{0em}{1pt}{1pt}
    \cline{2-6}
    \specialrule{0em}{1pt}{1pt}
    & NID-SVM                    & 99.61 $\pm$ 0.12   & 85.10 $\pm$ 4.34   & 94.11 $\pm$ 0.18   & 89.32 $\pm$ 2.73   \\
    & NID-\textbf{DA}-SVM        & \textbf{99.73} $\pm$ \textbf{0.06}   & \textbf{90.03} $\pm$ \textbf{4.12}   & \textbf{95.12} $\pm$ \textbf{1.04}   & \textbf{92.44} $\pm$ \textbf{1.65}   \\
    \specialrule{0em}{1pt}{1pt}
    \hline
    \specialrule{0em}{1pt}{1pt}
    \multirow{4}*{'saint'}
    & NID-LR                     & 98.22 $\pm$ 0.22   & 40.17 $\pm$ 2.89   & 96.47 $\pm$ 1.45   & 56.65 $\pm$ 2.81   \\
    & NID-\textbf{DA}-LR         & \textbf{98.47} $\pm$ \textbf{0.04}   & \textbf{43.78} $\pm$ \textbf{0.75}   & \textbf{96.93} $\pm$ \textbf{0.65}   & \textbf{60.32} $\pm$ \textbf{0.83}   \\
    \specialrule{0em}{1pt}{1pt}
    \cline{2-6}
    \specialrule{0em}{1pt}{1pt}
    & NID-SVM                    & 98.56 $\pm$ 0.03   & 45.39 $\pm$ 0.50   & 96.41 $\pm$ 0.75   & 61.72 $\pm$ 0.46   \\
    & NID-\textbf{DA}-SVM        & \textbf{98.60} $\pm$ \textbf{0.01}   & \textbf{46.08} $\pm$ \textbf{0.24}   & \textbf{97.47} $\pm$ \textbf{0.33}   & \textbf{62.58} $\pm$ \textbf{0.27}   \\
    \specialrule{0em}{1pt}{1pt}
    \hline
    \specialrule{0em}{1pt}{1pt}
    \multirow{4}*{'guesspasswd'}
    & NID-LR                     & 88.59 $\pm$ 0.75   & 34.21 $\pm$ 1.50   & 75.10 $\pm$ 2.87   & 46.98 $\pm$ 1.44   \\
    & NID-\textbf{DA}-LR         & \textbf{89.07} $\pm$ \textbf{0.24}   & \textbf{35.65} $\pm$ \textbf{0.59}   & \textbf{77.74} $\pm$ \textbf{0.19}   & \textbf{48.89} $\pm$ \textbf{0.57}   \\
    \specialrule{0em}{1pt}{1pt}
    \cline{2-6}
    \specialrule{0em}{1pt}{1pt}
    & NID-SVM                    & 94.59 $\pm$ 0.27   & 89.06 $\pm$ 1.93   & 22.21 $\pm$ 4.10   & 35.42 $\pm$ 5.19   \\
    & NID-\textbf{DA}-SVM        & \textbf{98.95} $\pm$ \textbf{0.10}   & \textbf{90.57} $\pm$ \textbf{2.17}   & \textbf{94.29} $\pm$ \textbf{0.22}   & \textbf{92.38} $\pm$ \textbf{1.19}   \\
    \specialrule{0em}{1pt}{1pt}
    \hline
    \specialrule{0em}{1pt}{1pt}
    \multirow{4}*{'snmpgetattack'}
    & NID-LR                     & \textbf{88.67} $\pm$ \textbf{0.00}   & 0.00               & 0.00               & -                  \\
    & NID-\textbf{DA}-LR         & 80.42 $\pm$ 0.58   & \textbf{36.61} $\pm$ \textbf{0.69}   & \textbf{99.43} $\pm$ \textbf{0.07}   & \textbf{53.51} $\pm$ \textbf{0.72}   \\
    \specialrule{0em}{1pt}{1pt}
    \cline{2-6}
    \specialrule{0em}{1pt}{1pt}
    & NID-SVM                    & \textbf{88.65} $\pm$ \textbf{0.02}   & 0.00               & 0.00               & -                  \\
    & NID-\textbf{DA}-SVM        & 82.42 $\pm$ 0.03   & \textbf{39.13} $\pm$ \textbf{0.03}   & \textbf{99.39} $\pm$ \textbf{0.21}   & \textbf{56.15} $\pm$ \textbf{0.04}   \\
    \specialrule{0em}{1pt}{1pt}
    \hline
    \specialrule{0em}{1pt}{1pt}
    \multirow{4}*{'snmpguess'}
    & NID-LR                     & 98.84 $\pm$ 0.15   & 78.61 $\pm$ 2.51   & \textbf{95.93} $\pm$ \textbf{0.05}   & 86.39 $\pm$ 1.55   \\
    & NID-\textbf{DA}-LR         & \textbf{99.07} $\pm$ \textbf{0.06}   & \textbf{82.66} $\pm$ \textbf{1.17}   & 95.84 $\pm$ 0.00   & \textbf{88.76} $\pm$ \textbf{0.68}   \\
    \specialrule{0em}{1pt}{1pt}
    \cline{2-6}
    \specialrule{0em}{1pt}{1pt}
    & NID-SVM                    & \textbf{96.18} $\pm$ \textbf{0.00}   & 0.00               & 0.00               & -                  \\
    & NID-\textbf{DA}-SVM        & 81.20 $\pm$ 0.04   & \textbf{16.85} $\pm$ \textbf{0.02}   & \textbf{99.72} $\pm$ \textbf{0.10}   & \textbf{28.83} $\pm$ \textbf{0.03}   \\
    \specialrule{0em}{1pt}{1pt}
    \hline
    \hline
    \end{tabular}
\end{table*}

In this subsection, we evaluate binary classification performance of the DA enhanced NID framework. In each group of the experiments, it is required to identify positive samples (i.e., one type of network intrusions) from negative samples (i.e., normal network requests).

As illustrated in Fig. \ref{Fig-NIDFramework}, a proportion of 6,000 randomly selected normal request data and 50 intrusion samples (augmented to 500 by the DA module) are used for training ML based NID models. In this case, LR and SVM are adopted since they are basic building blocks of many learning based IDSs \cite{16TOC-NIDvFeaSelect} \cite{18-NIDvDL}.

The training and testing procedures are repeated 15 times, each with varying selected negative samples and rejecting anomalous results. Afterwards, the statistic average of evaluation metrics are computed for validation. In this scenario, the adverse impact of data imbalance (i.e., training with all negative samples) is further reduced. Besides, it avoids training classifiers with biased data, such that the selected negative data are not representative.

Table \ref{Tab-Res2C-LRSVM} summarizes the binary classification results. It demonstrates that the DA enhanced NID frameworks (named as NID-DA-LR/SVM) achieve improved or comparable recall, and significantly outperforms other IDSs (named as NID-LR/SVM) in terms of precision and F1-score. Those results show that the intrusions augmented by the proposed DA module can be used to train potent classification models for NID tasks.

Note that the proposed frameworks obtain improved accuracy on most intrusion types except \verb"snmpgetattack" attack and \verb"snmpguess" attack of R2L. The reasons are two aspects. Firstly, computing accuracy requires to count normal network requests which occupy a large proportion in the testing set (see Table \ref{Tab-Data-KDDCup99} and \ref{Tab-Data-8Intrusions}). Thus, accuracy merely increases in small margin when additional intrusions are detected. Secondly, the low precision and recall of NID-LR/SVM on \verb"snmpgetattack" and \verb"snmpguess" attacks indicate that considerable intrusions are misclassified. Accordingly, they are not applicable to detect intrusions in the small sample scenario.

\subsection{Multiple Classification based NID}

\begin{table*}[t]
\centering
\setlength{\abovecaptionskip}{0pt}
\setlength{\belowcaptionskip}{0pt}
  \caption{ML based 4-class NID with SVM\protect\linebreak(Mean $\pm$ Std-Dev Percent)}
  \label{Tab-Res4C-LRSVM}
  \renewcommand\arraystretch{1.1}
  \centering
    \begin{tabular}{*{6}{p{1.5cm} p{2.0cm} c c c c}}
    \hline
    \hline
    \specialrule{0em}{2pt}{2pt}
    \multicolumn{1}{l}{{\textbf{Category}}} & \multicolumn{1}{l}{\textbf{NID Model Name}} & \multicolumn{1}{c}{\textbf{Accuracy}} & \multicolumn{1}{c}{\textbf{Precision}} & \multicolumn{1}{c}{\textbf{Recall}} & \multicolumn{1}{c}{\textbf{F1-Score}} \\
    \specialrule{0em}{2pt}{2pt}
    \hline
    \specialrule{0em}{1pt}{1pt}
    \multirow{3}*{\textbf{NORMAL}}
    & NID-SVM                       & 76.30 $\pm$ 0.01   & 75.91 $\pm$ 0.01   & \textbf{98.69} $\pm$ \textbf{0.02}   & 85.81 $\pm$ 0.01   \\
    & NID-\textbf{PGM}-SVM          & 76.16 $\pm$ 0.08   & 75.90 $\pm$ 0.03   & 98.41 $\pm$ 0.11   & 85.70 $\pm$ 0.06   \\
    & NID-\textbf{DA}-SVM    & \textbf{82.87} $\pm$ \textbf{0.17}   & \textbf{98.39} $\pm$ \textbf{0.20}   & 77.68 $\pm$ 0.32   & \textbf{86.82} $\pm$ \textbf{0.16}   \\
    \specialrule{0em}{1pt}{1pt}
    \hline
    \specialrule{0em}{1pt}{1pt}
    \multirow{3}*{\textbf{DOS}}
    & NID-SVM                       & 94.00 $\pm$ 0.01   & 93.35 $\pm$ 1.15   & 25.34 $\pm$ 0.16   & 39.86 $\pm$ 0.10   \\
    & NID-\textbf{PGM}-SVM          & 93.84 $\pm$ 0.10   & 86.98 $\pm$ 2.63   & 25.33 $\pm$ 0.49   & 39.23 $\pm$ 0.80   \\
    & NID-\textbf{DA}-SVM    & \textbf{99.48} $\pm$ \textbf{0.05}   & \textbf{94.09} $\pm$ \textbf{0.58}   & \textbf{99.70} $\pm$ \textbf{0.06}   & \textbf{96.81} $\pm$ \textbf{0.29}   \\
    \specialrule{0em}{1pt}{1pt}
    \hline
    \specialrule{0em}{1pt}{1pt}
    \multirow{3}*{\textbf{PROBE}}
    & NID-SVM                       & 98.82 $\pm$ 0.02   & 68.54 $\pm$ 0.43   & 83.27 $\pm$ 0.53   & 75.19 $\pm$ 0.37   \\
    & NID-\textbf{PGM}-SVM          & 98.80 $\pm$ 0.02   & 67.39 $\pm$ 0.31   & \textbf{85.89} $\pm$ \textbf{0.38}   & 75.53 $\pm$ 0.30   \\
    & NID-\textbf{DA}-SVM    & \textbf{99.32} $\pm$ \textbf{0.04}   & \textbf{88.70} $\pm$ \textbf{0.88}   & 78.13 $\pm$ 1.85   & \textbf{83.07} $\pm$ \textbf{1.00}   \\
    \specialrule{0em}{1pt}{1pt}
    \hline
    \specialrule{0em}{1pt}{1pt}
    \multirow{3}*{\textbf{R2L}}
    & NID-SVM                       & 83.43 $\pm$ 0.01   & \textbf{97.91} $\pm$ \textbf{0.64}   & 4.83 $\pm$ 0.02    & 9.20 $\pm$ 0.04    \\
    & NID-\textbf{PGM}-SVM          & 83.34 $\pm$ 0.05   & 94.15 $\pm$ 5.15   & 4.51 $\pm$ 0.02    & 8.60 $\pm$ 0.04    \\
    & NID-\textbf{DA}-SVM    & \textbf{83.74} $\pm$ \textbf{0.19}   & 51.75 $\pm$ 0.31   & \textbf{96.57} $\pm$ \textbf{0.66}   & \textbf{67.38} $\pm$ \textbf{0.23}   \\
    \specialrule{0em}{1pt}{1pt}
    \hline
    \hline
    \end{tabular}
\end{table*}

\begin{table*}[t]
\centering
\setlength{\abovecaptionskip}{0pt}
\setlength{\belowcaptionskip}{0pt}
  \caption{DL based 4-class NID with DNN\protect\linebreak(Mean $\pm$ Std-Dev Percent)}
  \label{Tab-Res4C-DNN}
  \renewcommand\arraystretch{1.1}
  \centering
    \begin{tabular}{*{6}{p{1.5cm} p{2.0cm} c c c c}}
    \hline
    \hline
    \specialrule{0em}{2pt}{2pt}
    \multicolumn{1}{l}{{\textbf{Category}}} & \multicolumn{1}{l}{\textbf{NID Model Name}} & \multicolumn{1}{c}{\textbf{Accuracy}} & \multicolumn{1}{c}{\textbf{Precision}} & \multicolumn{1}{c}{\textbf{Recall}} & \multicolumn{1}{c}{\textbf{F1-Score}} \\
    \specialrule{0em}{2pt}{2pt}
    \hline
    \specialrule{0em}{1pt}{1pt}
    \multirow{3}*{\textbf{NORMAL}}
    & NID-DNN                      & 77.15 $\pm$ 3.22   & 76.5 $\pm$ 2.81   & 99.15 $\pm$ 0.51   & 86.33 $\pm$ 1.64   \\
    & NID-\textbf{PGM}-DNN         & 83.17 $\pm$ 1.40   & 81.58 $\pm$ 1.31   & \textbf{99.26} $\pm$ \textbf{0.37}   & 89.55 $\pm$ 0.76   \\
    & NID-\textbf{DA}-DNN   & \textbf{87.96} $\pm$ \textbf{0.12}   & \textbf{86.95} $\pm$ \textbf{0.28}   & 98.15 $\pm$ 0.61   & \textbf{92.21} $\pm$ \textbf{0.11}   \\
    \specialrule{0em}{1pt}{1pt}
    \hline
    \specialrule{0em}{1pt}{1pt}
    \multirow{3}*{\textbf{DOS}}
    & NID-DNN                      & 93.81 $\pm$ 2.94   & 89.70 $\pm$ 6.58  & 23.55 $\pm$ 9.17  & 27.72 $\pm$ 7.37  \\
    & NID-\textbf{PGM}-DNN         & 99.04 $\pm$ 0.22   & \textbf{97.84} $\pm$ \textbf{3.73}   & 89.97 $\pm$ 0.98   & 93.7 $\pm$ 1.34   \\
    & NID-\textbf{DA}-DNN   & \textbf{99.59} $\pm$ \textbf{0.08}   & 97.14 $\pm$ 1.36   & \textbf{97.65} $\pm$ \textbf{0.40}   & \textbf{97.39} $\pm$ \textbf{0.52}   \\
    \specialrule{0em}{1pt}{1pt}
    \hline
    \specialrule{0em}{1pt}{1pt}
    \multirow{3}*{\textbf{PROBE}}
    & NID-DNN                      & 98.92 $\pm$ 0.31   & 74.54 $\pm$ 1.93   & 81.49 $\pm$ 8.26   & 76.79 $\pm$ 3.29   \\
    & NID-\textbf{PGM}-DNN         & 98.40 $\pm$ 0.07   & 64.93 $\pm$ 1.90   & 55.84 $\pm$ 7.72   & 59.78 $\pm$ 3.86   \\
    & NID-\textbf{DA}-DNN   & \textbf{99.05} $\pm$ \textbf{0.07}   & \textbf{75.27} $\pm$ \textbf{1.24}   & \textbf{83.47} $\pm$ \textbf{4.52}   & \textbf{79.11} $\pm$ \textbf{2.17}   \\
    \specialrule{0em}{1pt}{1pt}
    \hline
    \specialrule{0em}{1pt}{1pt}
    \multirow{3}*{\textbf{R2L}}
    & NID-DNN                      & 83.90 $\pm$ 2.72   & 58.99 $\pm$ 3.87   & 7.55 $\pm$ 5.81    & 11.28 $\pm$ 3.13   \\
    & NID-\textbf{PGM}-DNN         & 84.90 $\pm$ 1.26   & 88.45 $\pm$ 5.67   & 13.92 $\pm$ 7.48   & 23.60 $\pm$ 2.57   \\
    & NID-\textbf{DA}-DNN   & \textbf{89.09} $\pm$ \textbf{0.10}   & \textbf{92.17} $\pm$ \textbf{4.86}   & \textbf{40.97} $\pm$ \textbf{1.82}   & \textbf{56.64} $\pm$ \textbf{0.92}   \\
    \specialrule{0em}{1pt}{1pt}
    \hline
    \hline
    \end{tabular}
\end{table*}

In order to verify the performance of the DA module on enhancing the existing learning based IDSs, we have undertaken multi-class based NID experiments. In this case, the network data are categorized as: NORMAL, DOS, PROBE, and R2L, which covers the aforementioned 8 intrusion types.

For learning based IDSs, SVM and one typical DNN architecture presented in \cite{16-NIDvDL4SDN} \cite{17ICACCI-DNN4NID} are chosen as supervised NID models. In the training process of DNN, the batch size is initialized as 32 and the learning rate is adaptively decided. Furthermore, dropout \cite{Dropout} and cross-validation are employed to avoid over-fitting problems.

The evaluation stage is repeated 15 times. At each time, 6,000 and 12,000 normal requests are randomly selected for training SVM and DNN, respectively. In addition, 50 fixed samples of each type of intrusions are separately augmented to 500 by PGM and PGM-DGNNs associated DA module to 500. Overall, the augmented training set contains 4,000 intrusion samples. At the completion of training, the deviant results (e.g., extremely low F1-score) are rejected and the average metrics are computed for the follow-up analysis.

Table \ref{Tab-Res4C-LRSVM} and Table \ref{Tab-Res4C-DNN} present the comparison results of multi-class NID using SVM and DNN, respectively. It can be observed that the PGM-DGNNs enhanced NID frameworks (named as NID-DA-SVM/DNN) outperforms other two NID models with regard to both accuracy and F1-score.

For DNN based NID, the precision and recall obtained by the proposed frameworks are better than, or comparable to those obtained by other NID models. For SVM based NID, it achieves the highest precision and recall on DOS with notable exceptions on other three categories. The reason is that network intrusion data generated by $G$-net are strictly affected by DNN implemented $D$-net. As a result, augmented intrusion features are more sensitive for DNN based NID models.

Despite that PGM enhanced IDSs have achieved improvement on some evaluation metrics, those IDSs show undesired performance (e.g., low F1-score on PROBE when DNN is used) in identifying certain network intrusions. In contrast, PGM-DGNNs aided NID frameworks are able to accurately detect the intrusions with small and imbalanced samples.

\section{Discussion}
\label{Section-Discussion}

The DA enhanced NID framework proposed in this paper is trained in a data-driven manner. Its performance is closely related to the quality of augmented network intrusions with regard to the quantity and quality of known samples. Therefore, the selected network intrusion samples are recommended to be able to represent the underlying characteristics of relevant intrusion types.

In the proposed NID framework, we have presented the insight of using deep adversarial learning for augmenting limited network intrusion samples. The structure of $D$-net and $G$-net are generally formulated as fully connected DNNs, in which advanced network architectures \cite{W-GAN} might be employed to design DGNNs.

The computational complexity of the proposed NID framework includes the following two phases. In the testing phase, classifying over 80,000 network data takes a few seconds even if DNN is applied. In the training phase, time consumption can be analyzed as follows. First, PGM simulates synthesised intrusion data within minutes by using MCMC based Gibbs sampling. Second, DGNNs is efficiently optimized on high performance computing devices. In our experiments, training the DGNNs on NVIDIA GTX 1080 GPU takes less than two hours. Third, training supervised classifiers (i.e., SVM and DNN) requires less than half an hour. In spite of the training time, it demonstrates high efficiency in identifying network intrusions, which is necessary in network security.

The proposed DA module can be alternatively applied to assist NID algorithms implemented on distributed platforms. Specifically, the imbalanced training set is firstly augmented with the proposed DA module in a data center. The augmented dataset can then be partitioned into several data blocks, in which normal network request data and network intrusion samples have comparable proportions. Finally, those balanced data blocks are delivered to different computing nodes to distributively train NID models.

\section{Conclusion}
\label{Section-Conclusion}

In this paper, a general NID framework and two learning based data augmentation components have been jointly proposed to tackle the data scarcity and data imbalance problem in designing learning based IDSs. In this framework, statistic learning based PGM and deep learning based DGNNs of DA module are developed for enlarging limited intrusion samples in the training set. By employing classical ML models (e.g., LR and SVM) as well as advanced DNNs, it can accurately classify normal network request and heterogeneous network intrusions. Extensive experimental validations have been conducted on KDD Cup 99 dataset. Both binary and multiple classification results have shown that the DA enhanced IDSs outperform others regarding F1-score (which is a crucial criterion of evaluating imbalanced classification tasks). Additionally, it achieves improved or comparable accuracy, precision and recall, especially when DNN is adopted for classifying network data.



\bibliographystyle{IEEEbib}
\bibliography{refsAbrv}

\end{document}